\documentstyle[preprint,aps]{revtex}

\begin{document}
\draft

\title{DOES THE QUENCHED KARDAR-PARISI-ZHANG EQUATION DESCRIBE THE DIRECTED PERCOLATION
DEPINNING MODELS?}

\author{A. D\'{\i}az-S\'anchez \cite{mimail}}
\address{Departamento de F\'{\i}sica, Universidad de Murcia, E-30071
Murcia, Spain}

\author{L. A. Braunstein \cite{tumail} and R. C. Buceta}
\address{Departamento de F\'{\i}sica, Facultad de Ciencias
Exactas y Naturales,\\ Universidad Nacional de Mar del Plata,
Argentina}

\date{\today}
\maketitle

\begin{abstract}

The roughening of interfaces moving in inhomogeneous media is
investigated by numerical integration of the phenomenological
stochastic differential equation proposed by Kardar, Parisi, and
Zhang [Phys. Rev. Lett. {\bf 56}, 889, (1986)] with quenched noise
(QKPZ). We express the evolution equations for the mean height and
the roughness into two contributions: the local and the lateral
one. We compare this two contributions with the ones obtained for
two directed percolation deppining models (DPD): the Tang and
Leschhorn model [Phys. Rev A {\bf 45}, R8309 (1992)] and the
Buldyrev {\sl et al.} model [Phys. Rev. A {\bf 45}, R8313 (1992)]
by Braunstein {\sl et al.} [J. Phys. A {\bf 32}, 1801 (1999);
Phys. Rev. E {\bf 59}, 4243 (1999)]. Even these models have being
classified in the same universality class that the QKPZ the
contributions to the growing mechanisms are quite different. The
lateral contribution in the DPD models, leads to an increasing of
the roughness near the criticality while in the QKPZ  equation
this contribution always flattens the roughness. These results
suggest that the QKPZ equation does not describe properly the DPD
models even when the exponents derived from this equation are
similar to the one obtained from simulations of these models.

\end{abstract}


\section{INTRODUCTION}
\label{sec:intro}

The description of the noise-driven growth that leads to
self-affine interface far from equilibrium is a challenging
problem. The interface has been characterized through scaling of
the interfacial width $w$ with time $t$ and lateral size $L$. The
result is the determination of two exponents $\beta$ and $\alpha$
called dynamical and roughness exponents, respectively. It is well
known that interfacial width $w\sim L^\alpha$ for $t\gg t^*$ and
$w\sim t^\beta$ for $t\ll t^*$, where $t^*\simeq L^{\alpha/\beta}$
is the saturation time. These properties occur in many
experimental situations and many models of surface growth. The
values of the exponents leads to their classification in different
universality classes. Several models, belonging to the same
directed percolation depinning (DPD) universality class, have been
introduced to explain experiments on fluid imbibition in porous
media, roughening in slow combustion of paper, growth of bacterial
colonies, etc. It is currently accepted that the quenched disorder
plays an essential role in those experiments. The DPD models take
into account the most important features of the experiments
\cite{Buld,Horv}. On the other hand the phenomenological
stochastic differential equation proposed by Kardar, Parisi, and
Zhang \cite{kpz} with quenched noise (QKPZ) is used to describe
the interface growth in disordered media and drives to the same
universality class than the DPD models, in the sense that they
have the same exponents. Moreover, processes with the same
exponent may not belong to the same universality class. For
example, $1+1$-dimensional lattice gas simulations of roughening
of immiscible fluid-fluid interface \cite{Flekkoy} lead to the
same exponents as the $1+1$-dimensional KPZ \cite{kpz}
($\beta=1/3$ and $\alpha=1/2$) for surface growth, but this model
is completely linear, so there is no obvious mathematical
relationship between these two processes.

The two first DPD models were simultaneously introduced by
Buldyrev {\sl et al.} \cite{Buld} and Tang and Leschhorn
\cite{Tang} to explain the fluid imbibition in paper sheet.
Several authors have been focused their attention on scaling
properties and relationships between the dynamical and the static
exponent for these models. These two models have been recently
reviewed by Braunstein {\sl et al.} \cite{Brauns2,Brauns3} from a
different point of view than the traditional one. The principal
contribution was the restatement of the Microscopic Equation for
each model. These equations allows the separation into two
contributions: the local and the lateral (or contact) one. They
found that the lateral contribution to the temporal derivative of
the interface width (DSIW) may be either negative or positive and
that the behavior of this contribution depends on the pressure
$p$, where $p$ is the microscopic driving force. The negative
contribution tends to smooth out the surface, this case dominate
for $p \gg p_c$ (where $p_c$ is the critical pressure). The
positive contribution enhances the roughness. At the critical
pressure the local contribution to the DSIW is practically
constant, but the lateral contribution is very strong. This last
contribution, has important duties on the power law behavior in
the DPD models.

In this paper we focus the attention in the QKPZ equation in order
to compare the different contributions with the ones obtained for
the DPD models. In this context we show that the results obtained
from this equation are quite different from the ones obtained in
the DPD models. We separate the QKPZ equation into two
contributions: the local contribution and the lateral one. In this
context we study the mean height speed (MHS) and the DSIW as a
function of time. The paper is organized as follows. In
Section~\ref{sec:ecu} we separate the QKPZ equation into two
contributions for the mean height and the roughness. We study the
MHS, analyzing the local and the lateral contributions. Also, the
two contributions to the DSIW are analyzed. In
Section~\ref{sec:comp} we compare the DPD models with the QKPZ
model. Finally, we present the main conclusions in
Section~\ref{sec:concl}.

\section{MACROSCOPIC CONTRIBUTIONS FROM THE QKPZ EQUATION}
\label{sec:ecu}

\subsection{Equations}

The QKPZ equation for the surface height $h=h(x,t)$, in
$1+1$-dimension, is given by
\begin{equation}
\partial_t h={\cal{F}} + \nu \;\partial^2_x h + \lambda\;
(\partial_x h)^2 +\eta(x,h)\label{qkpz}
\end{equation}
where $\nu$ and $\lambda$ are constants and the noise depends on
position x and height $h$ with the properties that $<\eta(x,h)>=0$
and $<\eta(x,h)\;\eta(x',h')>=2D\;\delta(x-x')\;\delta(h-h')$
where $2D$ is the noise intensity.

We can distinguish two contributions to this equation, the local
growth ${\cal{S}}= {\cal{F}} +\eta(x,h)$ and the lateral one
${\cal{L}}=\nu\;\partial^2_x h + \lambda\; (\partial_x h)^2$. So
we can write the evolution equation for the mean height as
\begin{equation}
\partial_t h={\cal{S}} +{\cal{L}}\;.
\end{equation}
Taking the derivative of the square interface width,
$w^2=\langle (h-\langle h \rangle)^2\rangle$, its evolution equation
is given by
\begin{equation}
{\partial_t w}^2=2\langle (h-\langle h \rangle){\partial_t h}\rangle=
 2\langle (h-\langle h \rangle){\cal{S}}\rangle+
2\langle (h-\langle h \rangle){\cal{L}}\rangle
\end{equation}
where $\langle \dots \rangle$ means average over the lattice. The
first term can be identified as the local growth contribution, and
the second term as the lateral growth contribution. The separation
into these two analytical terms allows us to compare the
mechanisms of growth in the QKPZ equation with ones of the DPD
models. In the present paper we focus only on the dynamical
behavior, {\sl i.e.} $t \ll t^* \simeq L $ (in these models
$\alpha \sim \beta$) for the mean height and roughness.

We have performed the direct numerical integration of
Eq.~(\ref{qkpz}) in one dimension in the discretized version
\cite{Csahok,Jeong}
\begin{eqnarray*}
h(x,t+\triangle t)&=&h(x,t)+ \triangle
t\;\left\{\;h(x-1,t)+h(x+1,t)-2h(x,t)\right.\\&
&+\frac{\lambda}{8}\;\left\{h(x+1,t)-h(x-1,t)\right\}^2 \\ & &
+{\cal{F}}+\eta(x,[h(x,t)])\left. \right\}\;,
\end{eqnarray*}
where $[\dots]$ denotes the integer part and $\eta$ is uniformly
distributed in $[-a/2,a/2]$, where $a=10^{2/3}$ is selected. We
use $L=8192$ and $\triangle t =0.01$. The initial condition is
$h(x,0)=0$ and periodic boundary conditions are used. The averages
was taken over $100$ samples. We study this equation for different
values of $\lambda$ with $\nu=1$.

\subsection{Mean height}

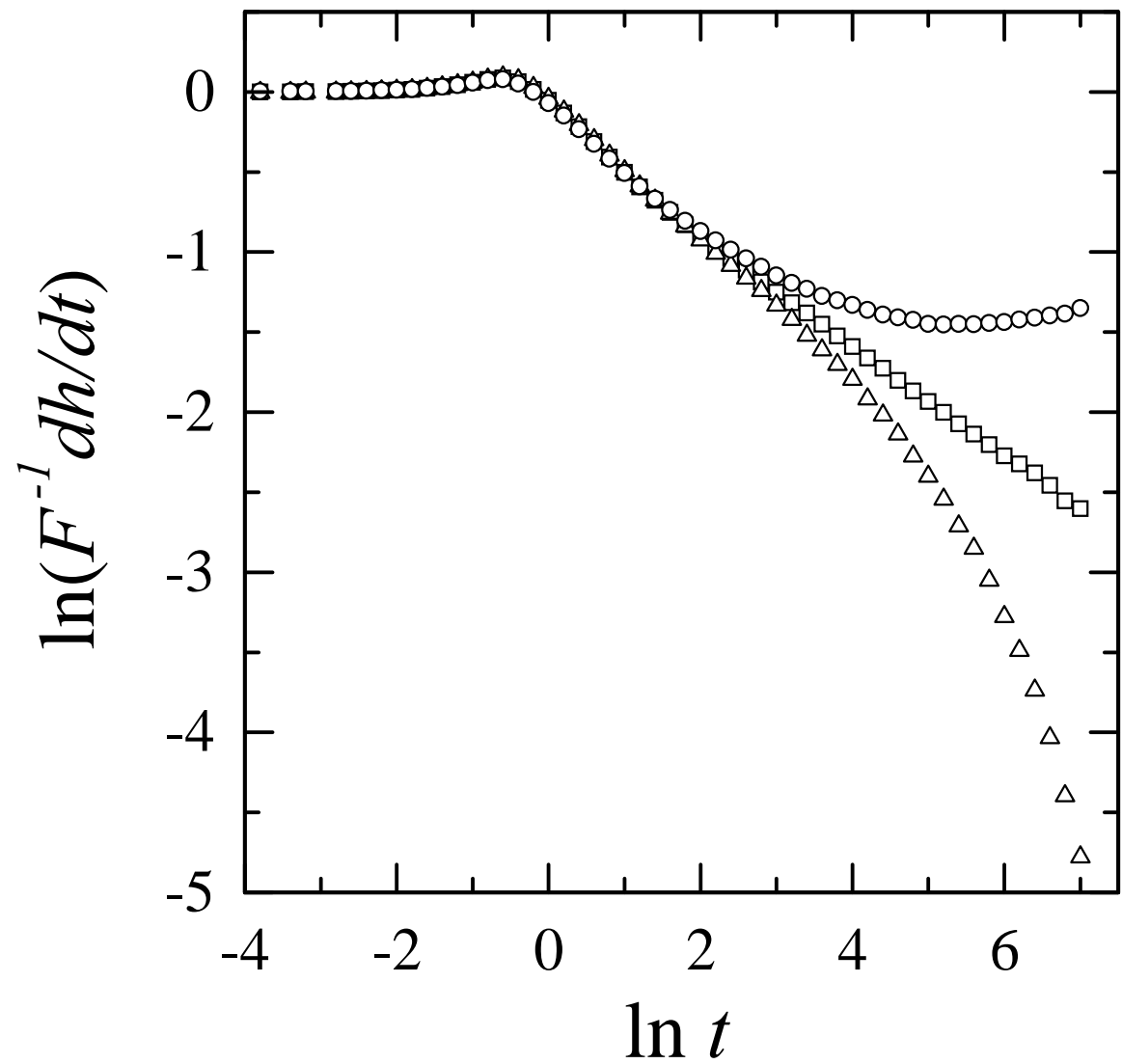
\begin{figure}
\caption{Plots of ${\cal{F}}^{-1}\,dh/dt$ vs $t$ for $\lambda=1$.
The parameter ${\cal{F}}$ is 0.51 ($\bigcirc$), 0.464 ($\Box$), and
0.43 ($\bigtriangleup$).} \label{dhdt1}
\end{figure}

In Fig.~\ref{dhdt1} we show the MHS as a function of time in the
three regimes (moving, critical and pinning phases) for
$\lambda=1$. The initial condition for the MHS is ${\cal{F}}$ in
all regimes. At the criticality we found for the mean height a
power law behavior with approximately the same dynamical exponent
that the roughness one, $\beta=0.67\pm 0.01$ for
${\cal{F}}_c=\;0.464$, where ${\cal{F}}_c$ is the critical driving
force. In the moving and pinning phases we can see that this power
law does not hold. Bellow the criticality, in the pinning phase,
the MHS goes to zero. In the moving phase (${\cal{F}} >
{\cal{F}}_c$), the MHS goes to certain constant value.

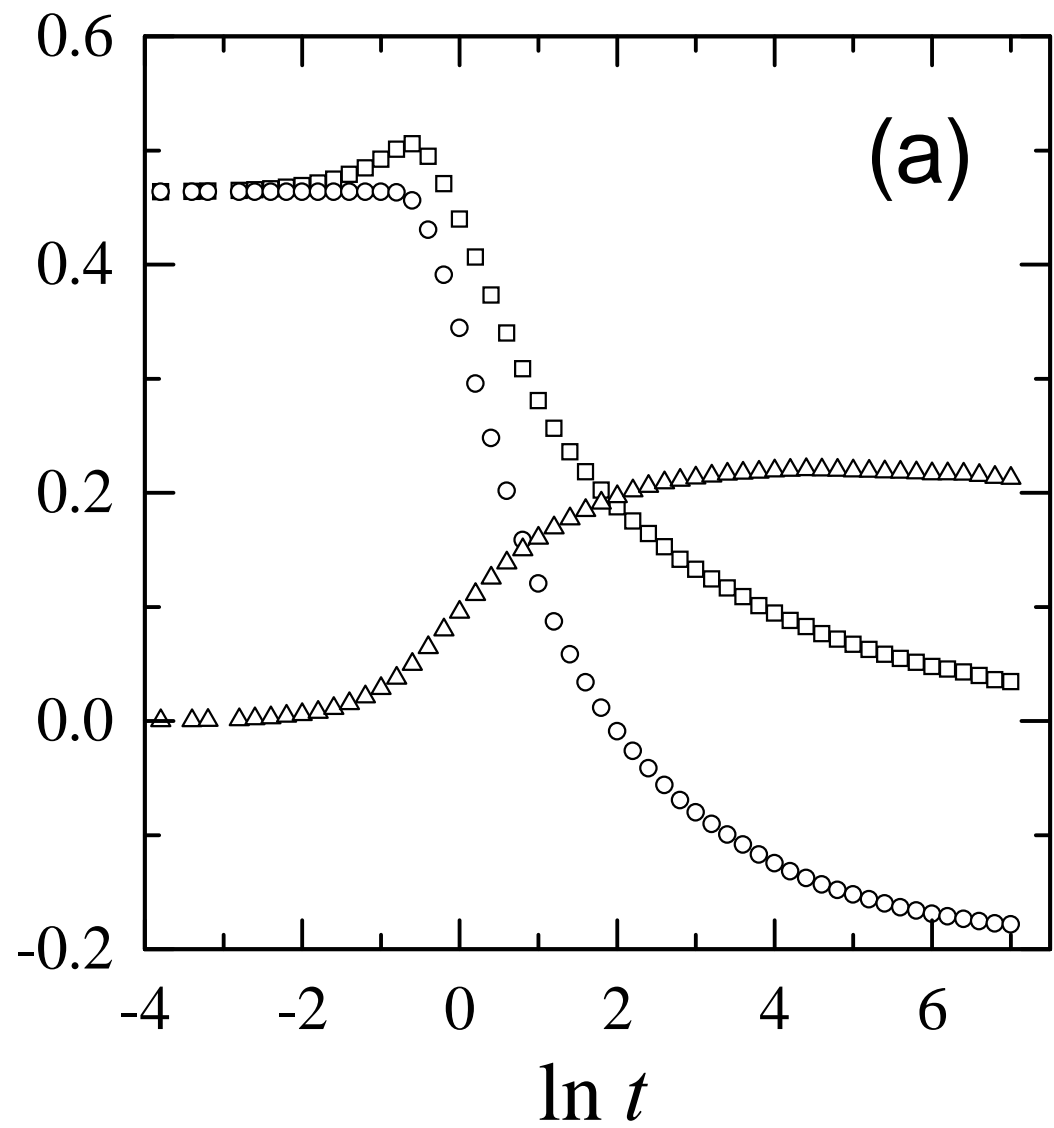
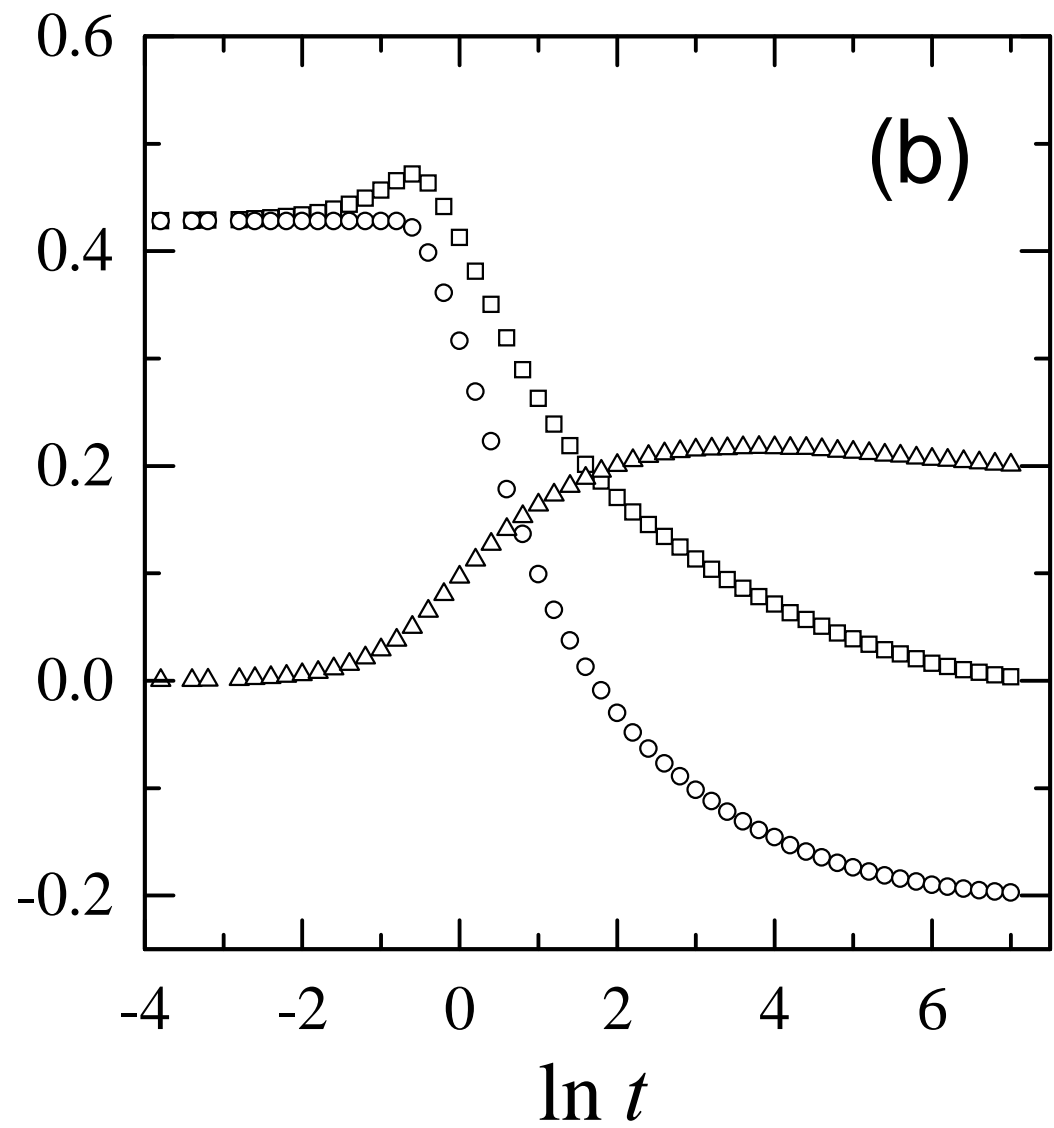
\begin{figure}
\caption{Semi-ln plots of the different contributions to the MHS
as a function of time for different values of ${\cal{F}}$ and
$\lambda=1$. The circles ($\bigcirc$) represent the local
contribution, the triangles ($\bigtriangleup$) represent the
lateral contribution, and the squares ($\Box$) represent the total
MHS. The (a) plot shows the critical phase ${\cal{F}}=0.464$. The
(b) plot shows the pinning phase ${\cal{F}}=0.43$. The (c) plot
shows the moving phase ${\cal{F}}=0.51$.} \label{dhdt2}
\end{figure}

In Fig.~\ref{dhdt2} we show the contributions to the MHS: the
local one $\langle{\cal{S}}\rangle$ and the lateral one
$\langle{\cal{L}}\rangle$. The local contribution, which is equal
to ${\cal{F}}$ at $t=0$, is stronger in the early time regime.
This is because in this regime the difference of heights between
nearest neighbors is very low and the contribution of the lateral
term is negligible. We see that the local contribution takes
negative values from $t \gtrsim 7$, so in this regime the local
contribution brakes the growing of $\langle h \rangle$. The
behavior of both contributions, lateral and local one, is not very
different in every phases although their sum only drive to a power
law at the criticality. In every phases both contributions are
equal at $t\simeq 2.5$. This means that at this time both
mechanisms are equal independently of ${\cal{F}}$. Increasing
$\lambda$ the lateral contribution is enhanced at shorter time,
for $\lambda=2$ the crossover is at $t\simeq 1$. In the asymptotic
dynamic regime both contributions are important and neither
dominates over the other one.

\subsection{Roughness}
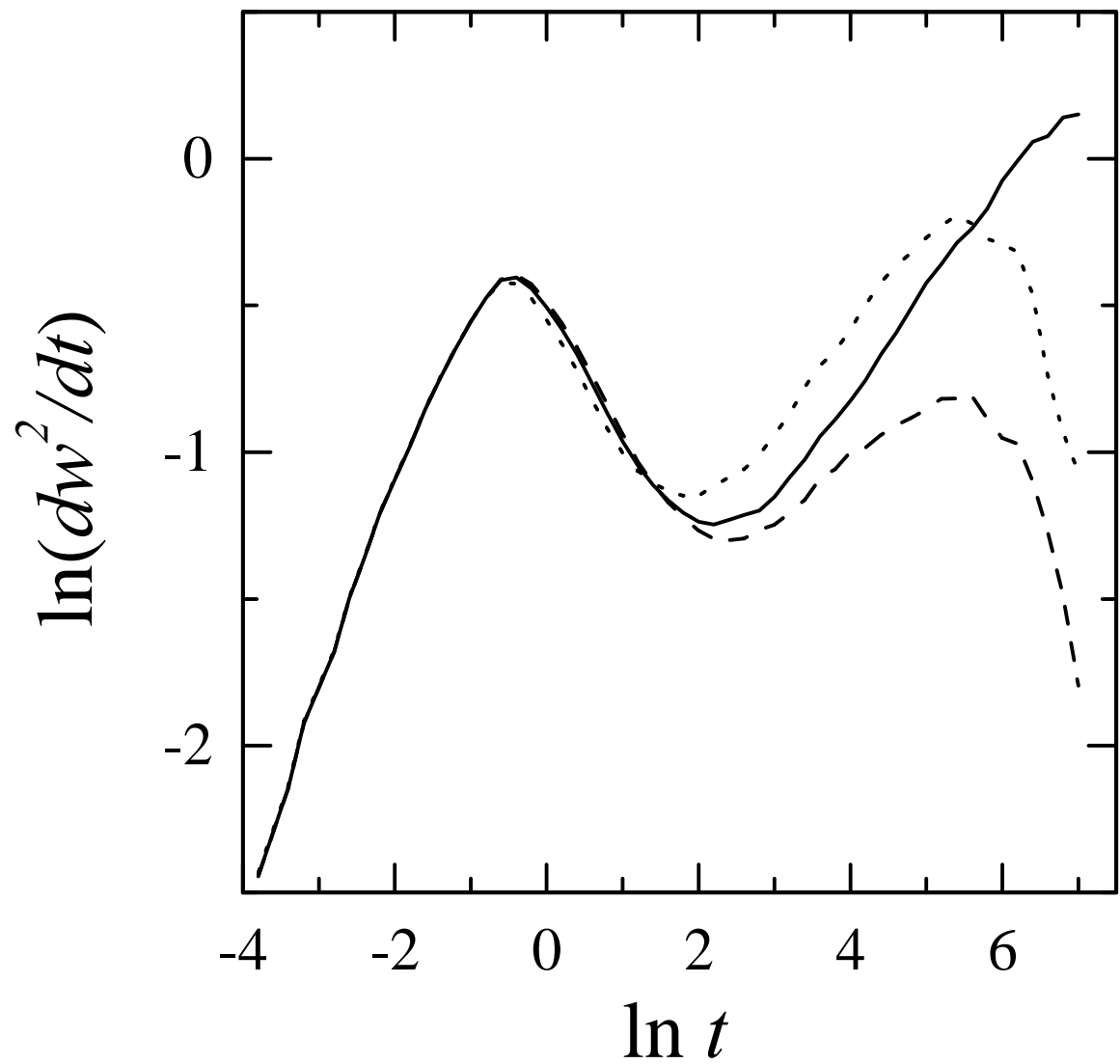
\begin{figure}
\caption{DSIW as a function of time in the critical, pinning, and
moving phases for $\lambda=1$. The parameter ${\cal{F}}$ is 0.464
(solid line), 0.43 (dashed line), and 0.54 (doted line).}
\label{dw2dt}
\end{figure}

Fig.~\ref{dw2dt} shows the temporal DSIW as a function of time for
various values of ${\cal{F}}$ and $\lambda=1$. Here we found that
the DSIW increases continuously from zero to a maximum value. The
power law holds only at the criticality. The DSIW goes
asymptotically to zero at the pinning and moving phases. The
dynamical exponent obtained for the roughness was $\beta=0.66 \pm
0.02$.

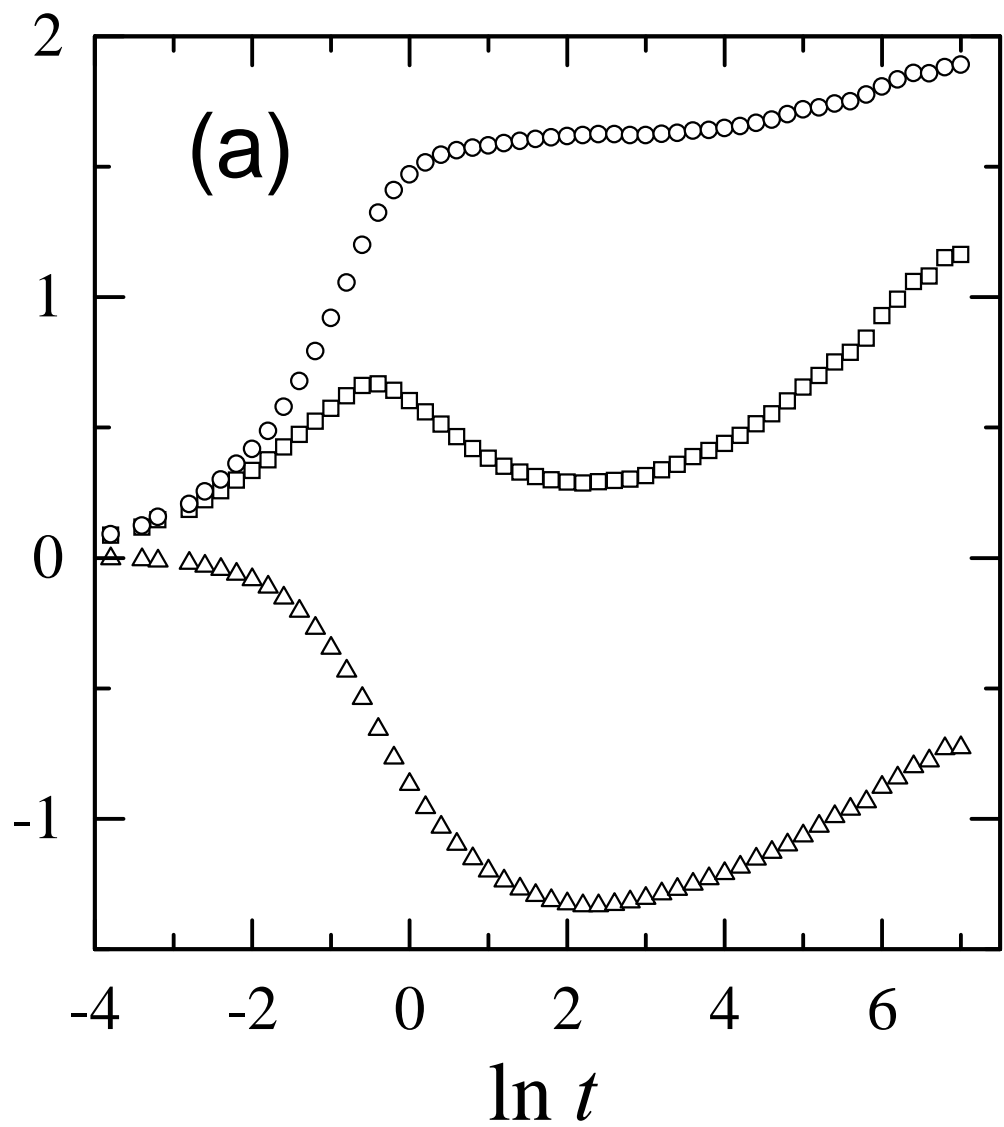
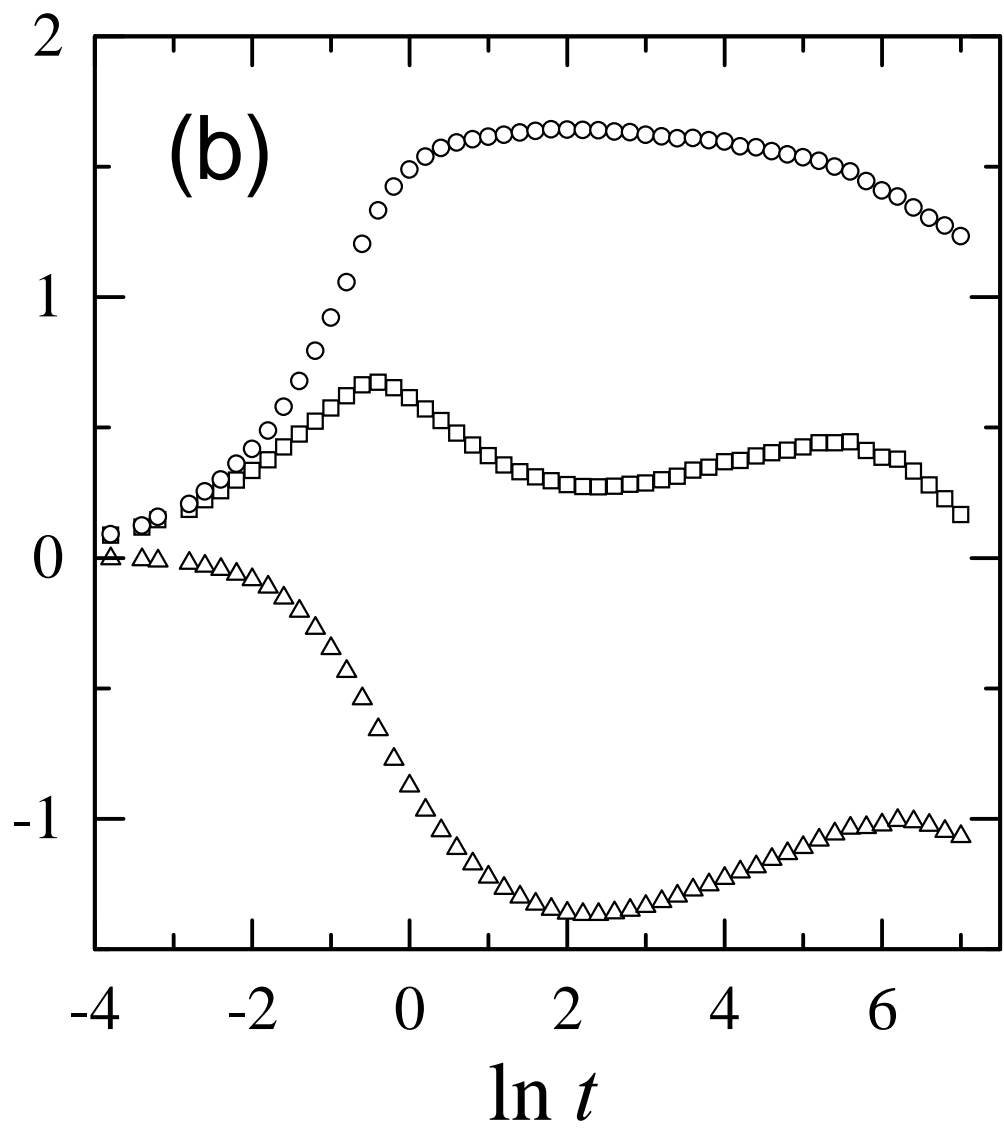
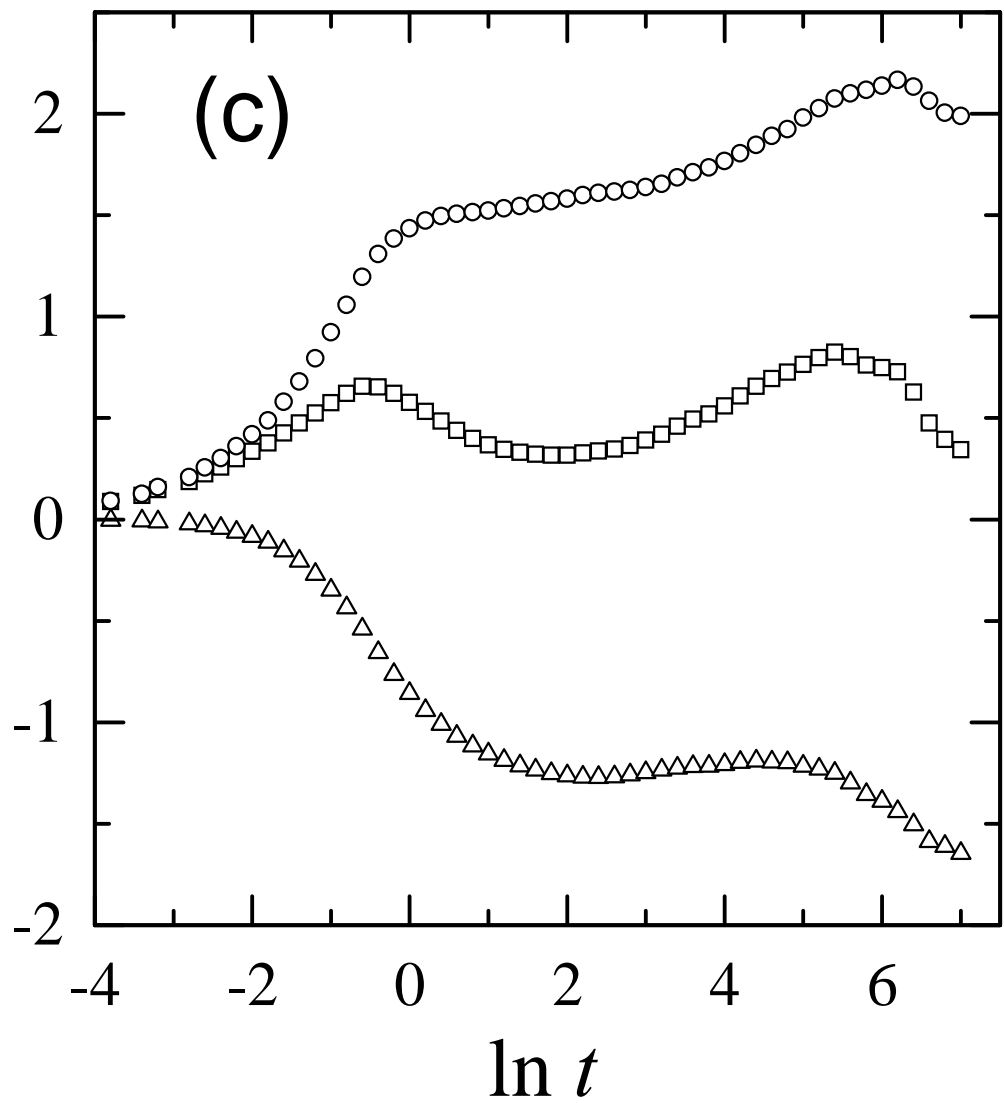
\begin{figure}
\caption{Semi-ln plots of the different contributions to the DSIW
as a function of time for different values of ${\cal{F}}$ and
$\lambda=1$. The circles ($\bigcirc$) represent the local
contribution, the triangles ($\bigtriangleup$) represent the
lateral contribution, and the squares ($\Box$) represent the total
DSIW. The (a) plot shows the critical phase ${\cal{F}}=0.464$. The
(b) plot shows the pinning phase ${\cal{F}}=0.43$. The (c) plot
shows the moving phase ${\cal{F}}=0.54$.}
 \label{dw2dt2}
\end{figure}

In Fig.~\ref{dw2dt2}, we show the two contributions to the DSIW
for different values of ${\cal{F}}$. The local contribution
$2\langle (h-\langle h \rangle){\cal{S}}\rangle$ to the DSIW is
always positive. As ${\cal{F}}$ decreases, this contribution also
decreases slowly, but always roughen the interface. On the other
hand, the lateral contribution $2\langle (h-\langle h
\rangle){\cal{L}}\rangle$ takes negative values in every phases,
smoothing out the surface.

In Fig.~\ref{dw2dt2lamb}, we plot the DSIW for three values of
$\lambda$ at the criticality. The slope $\beta$ is independent of
$\lambda$ although we found some differences in these plots. The
scaling dynamical regime is reached before at greater values of
$\lambda$. This is due to the fact that the lateral contribution,
which is the main responsible of the generation of correlations,
becomes more important earlier for larger values of $\lambda$.

\section{COMPARISONS WITH THE DPD MODELS}
\label{sec:comp}

In this section we present the similarities and differences
between the DPD models and the QKPZ equation. In previous work
Braunstein {\sl et al.} \cite{Brauns2,Brauns3} wrote the
microscopic equation for the TL and Buldyrev models. They
identified two separate contributions for the MHS and the DSIW:
the local and the lateral (contact) one. In the present paper, we
have also separate the QKPZ equation into these two contributions.

We found that, in the asymptotic dynamical regime, the behavior of
the MHS are similar in the QKPZ equation and in the DPD models in
every phases \cite{Brauns2,Brauns3}. At short time we found a
maximum value which depend on $\lambda$. This maximum value is not
found in the DPD models. Differences at short time are expected
because the QKPZ only describes the scaling limit or hydrodynamic
limit. The lateral and local contributions to the MHS in the DPD
models are qualitative and quantitative different from the QKPZ
ones. At long time the local contribution brakes the growing of
$\langle h \rangle$ in the QKPZ equation (in the DPD models this
contribution is always positive). Also, the contributions of the
QKPZ equation are not so much different in each phase, while in
the DPD models both contributions are very different behavior in
each phase.

In the asymptotic regime, the behavior of the DSIW is similar in
the QKPZ equation and in the DPD models. In the DPD models $p$ is
the initial condition in all regimes, this is due to the fact that
in the early regime the dynamics is like random deposition with
probability $p$ \cite{Brauns4,Lopez}. In the QKPZ equation the
DSIW increases continuously from zero to a maximum value, the
macroscopic equation presented by Braunstein {\sl et al.}
\cite{Brauns4,Brauns1} for the DPD models holds in the scaling
limit or hydrodynamic limit, but breaks down at short times as was
expected. The lateral and the local contributions to the DSIW is
quite different in each model. In the QKPZ  the lateral
contribution is always negative. This is the main difference with
the DPD models where for $p \lesssim p_c$, the lateral
contribution is always positive roughening the interface
\cite{Brauns2,Brauns3}. In the DPD models the lateral contribution
is enhanced by local growth, the lateral growth may also increase
the probability of local growth. This crossing interaction
mechanism makes the lateral growth dominant near the criticality
and this is due to the fact that the quenched noise is coupled to
the dynamic of the interface \cite{Brauns5}. In the QKPZ model
this cross mechanism between contributions is not taken into
account because the noise is additive. Moreover, the model are
said to belong to the same universality class that the QKPZ
equation because the exponents are quantitatively similar. Even
thought the behaviors of the contributions to the growth are
qualitatively and quantitatively different.

In the experiments the advancement of the interface is
determinated  by the coupled effect of the random distribution of
the capillary sizes, the surface tension and the local properties
of the flow, so it is not surprising that all these effect give
rise to a multiplicative noise. This multiplicative noise must be
taken into account at the time to pose a model with the essential
features of the experiment of surface growth in disordered media.
In the TL and the Buldyrev models the growing rules for the
evolution of the local height are strongly coupled to the quenched
noise in a multiplicative way. In both models the microscopic
rules that allows the growth from an unblocked cell
\cite{Brauns2,Brauns3} depends in some way on the local slope. In
that sense this coupled effect is not taken into account in the
QKPZ equation. The effect of a multiplicative noise as being
proposed by Csah\'ok {\sl et al.} \cite{Csahok} by means of a
phenomenological equation. They found a crossover between two
temporal regimes with $\beta=0.65$ to $\beta=0.26$ but the value
of $\alpha \simeq 0.47$ was obtained over a short range spatial
scale. Braunstein {\sl et al.} derived the continuous equation for
the TL model \cite{Brauns5}. This equation as the same term that
the QKPZ equation but its coefficients depends on the competition
between the driving force and the quenched noise. In that sense
the noise is multiplicative. This results joint to the results
obtained in this work supports that the QKPZ does not describe
fully the DPD models even if the exponents are quantitatively
similar.

\section{CONCLUSIONS}
\label{sec:concl}

We express the evolution equations of the QKPZ model for the mean
height and the roughness into two contributions: the local and the
lateral one. We found that the contributions to the growing
mechanisms are quite different from the DPD models. In the scaling
regime, the local contribution to the MHS brakes the growing of
$\langle h \rangle$ in the QKPZ model. The lateral contribution to
the DSIW is negative in all phases for the QKPZ model smoothing
out the interface. In the DPD models the lateral contribution is
always positive for $p \lesssim p_c$ roughening the interface.
Nevertheless, the DSIW and MHS gives the same scaling exponents in
these models, moreover it is not clear why. Our results suggest
that the QKPZ equation does not describe properly the dynamics of
the DPD models even if the exponents are similar.

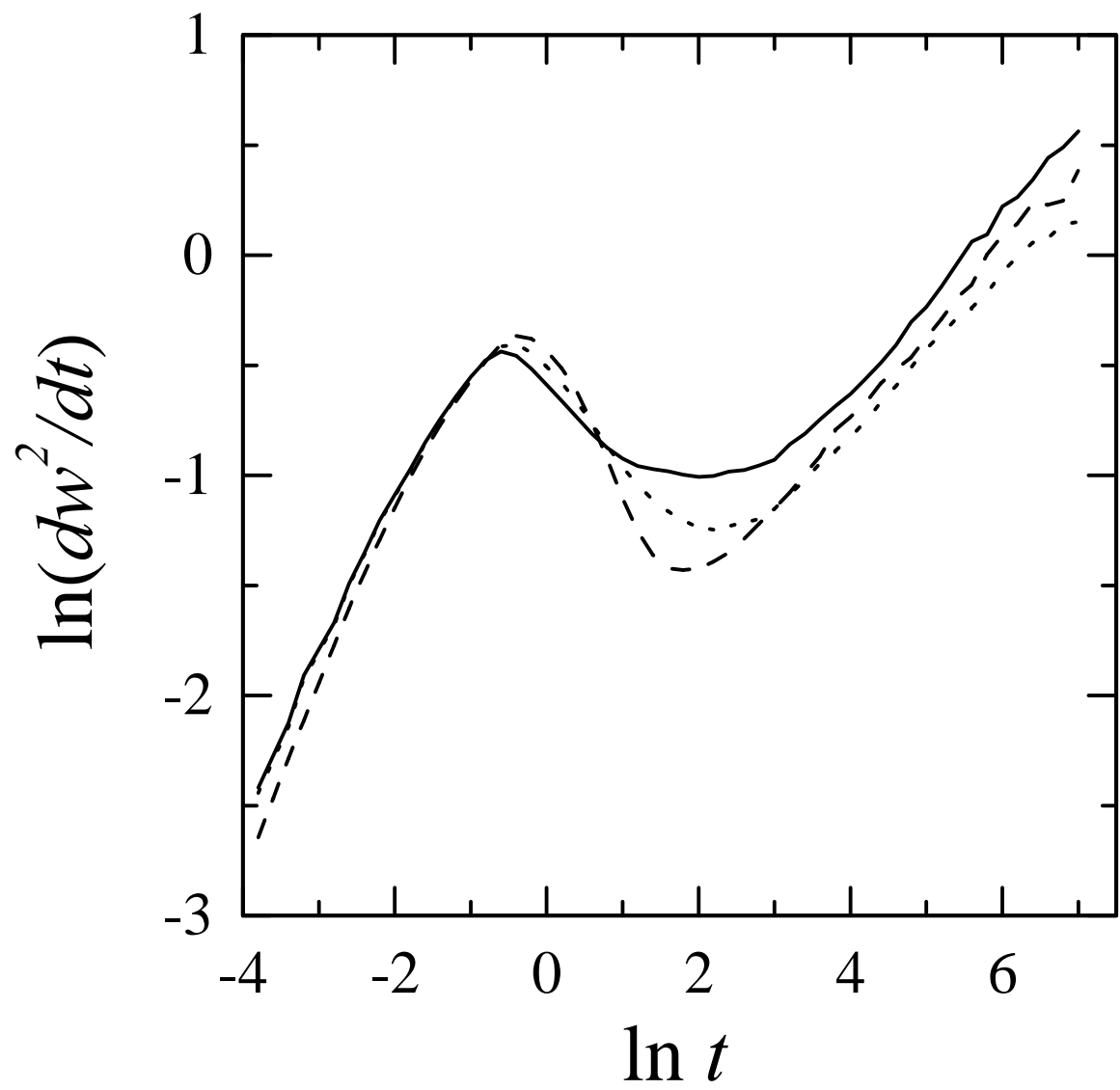
\begin{figure}
\caption{DSIW as a function of time in the critical regime for
different values of $\lambda$: 0.5 (solid line), 1 (doted line),
and 2 (dashed line) } \label{dw2dt2lamb}
\end{figure}
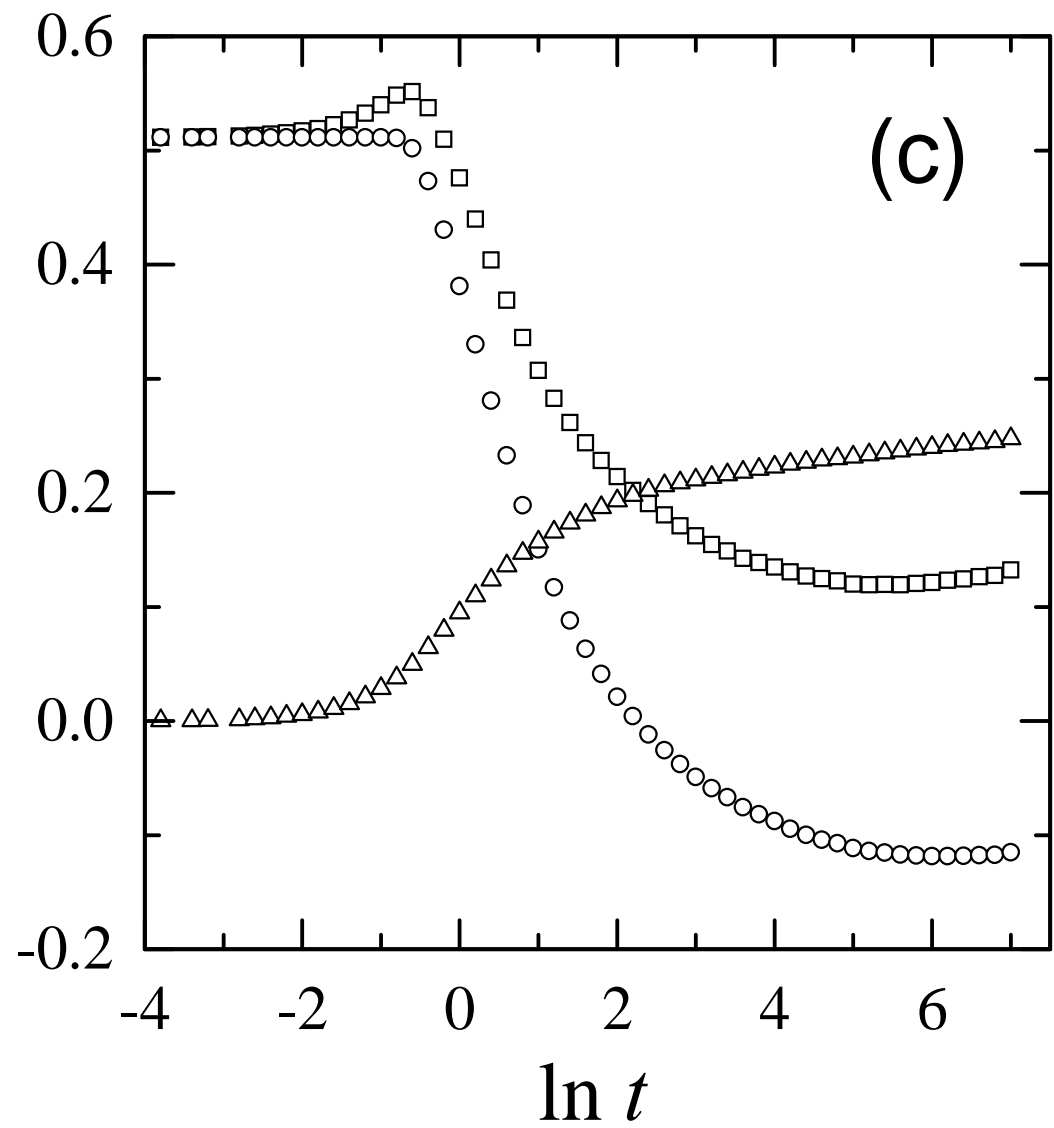
\end{document}